# Cobalt spin state above the valence and spin-state transition in $(Pr_{0.7}Sm_{0.3})_{0.7}Ca_{0.3}CoO_3$

F. Guillou, Y. Bréard and V. Hardy

*Laboratoire CRISMAT, ENSICAEN, UMR 6508 CNRS, 6 Boulevard du Maréchal Juin, 14050 Caen Cedex, France.*

**Abstract**

$(Pr_{0.7}Sm_{0.3})_{0.7}Ca_{0.3}CoO_3$ belongs to a class of cobalt oxides undergoing a first-order transition ($T^* \approx 90$ K) associated to a coupled change in the valence and spin-state degrees of freedom. The Curie-Weiss regime present around room temperature ($T >> T^*$) was analysed in detail to address the controversial issue of the cobalt spin states above the transition. This magnetic investigation indicates that the $Co^{4+}$ are in an intermediate spin-state, while the $Co^{3+}$ are in a mixed state combining low-spin and high-spin states. These results are discussed with respect to the literature on related compounds and recent results of x-ray absorption spectroscopy.

## 1. Introduction

Magnetic oxides containing cobalt are known to exhibit a great richness of physical properties. This can be ascribed to the fact that —besides the usual interplay between lattice, spins and charge carriers generally encountered in oxides— the presence of $Co^{3+}$ or $Co^{4+}$ introduces an additional degree of freedom associated to the nature of their spin states. The best known example is that of $Co^{3+}$ in an octahedral environment, for which the competition between low-spin (LS, $t_{2g}^6 e_g^0$), intermediate-spin (IS, $t_{2g}^5 e_g^1$) and high-spin (HS, $t_{2g}^4 e_g^2$) states is particularly subtle, leading to a situation prone to the development of spin state transitions (SST). The archetypical compound illustrating such a behavior is the perovskite $LaCoO_3$ which has been intensively studied since the 1950s [1]. As the temperature is increased, this compound undergoes two smooth transitions centered around $T_1 \sim 100$ K and $T_2 \sim 450$ K. While the $Co^{3+}$ spin state is unanimously recognized to be LS below $T_1$, the nature of the "higher" spin-states in the ranges $T_1 < T < T_2$ and $T > T_2$ is still very controversial (see for instance the overviews given in [2] and references therein).

Recently, another type of SST was revealed in orthorhombic perovskites of compositions $(Pr_{1-y}Ln_y)_{1-x}Ca_xCoO_3$, where Ln = Sm, Eu, Gd, Tb or Y, while $x$ and $y$ are approximately within the ranges 0.2-0.5 and 0-0.3, respectively [3-9]. Contrary to the crossover-like behavior observed in LaCoO$_3$, these compounds exhibit a sharp transition (at a temperature hereafter referred to as $T^*$) which can be regarded as being first-order in that it corresponds to abrupt jumps in magnetization, unit-cell volume and entropy [3,4,10]. Starting from 2010, there was an accumulation of results showing that a variation in the valence states of Pr and Co takes place along with the SST [9-12]. On the basis of x-ray absorption spectroscopy (XAS), it was demonstrated that a fraction of Pr is stabilized in a tetravalent state at low-$T$, and, when crossing $T^*$ upon warming, there is a $Pr^{4+}$-to-$Pr^{3+}$ transition counterbalanced by a corresponding change from $Co^{3+}$ to $Co^{4+}$ [12-14]. The transition in these compounds should thus be regarded as a coupled valence and spin-state transition (VSST).

The evolution of the cobalt spin states along the VSST is a controversial issue, bearing similarities with the situation encountered in LaCoO$_3$. While it is widely admitted that $Co^{3+}$ is LS below $T^*$, the nature of the spin state above $T^*$ remains an open question. Like for LaCoO$_3$ above $T_1$, the two main scenarios in competition about the $Co^{3+}$ spin-state above the VSST are either (i) the realization of the IS state, or (ii) the occurrence of an inhomogeneous mixed state involving LS and HS states. Although the former possibility was adopted in all the first studies on the VSST, recent XAS experiments rather lend support to the relevance of the LS/HS scenario [15,16]. The aim of the present study is to perform a quantitative analysis of the magnetic susceptibility at $T > T^*$ to shed new light on the nature of the Co spin state above the VSST. This analysis will be based on the approximation of a simple ionic approach, as done in most of the literature on cobalt oxides.

Our study is carried out on the compound $(Pr_{0.7}Sm_{0.3})_{0.7}Ca_{0.3}CoO_3$, which exhibits a transition at $T^* \approx 90$ K and is well documented in the literature [6,17]. As previously demonstrated in related compounds, one can consider that Pr is purely trivalent around room temperature ($T \gg T^*$) [5,6,10,14], which implies that the contents of $Co^{3+}$ and $Co^{4+}$ are equal to 0.7 and 0.3 per f.u., respectively. The two competing scenarios about the spin state of $Co^{3+}$ can thus be accounted for by writing the cationic distribution ($T \gg T^*$) as $(Pr^{3+})_{0.49}(Sm^{3+})_{0.21}(Ca^{2+})_{0.3}(Co^{4+})_{0.3}(Co^{3+}LS)_{0.7-z}(Co^{3+} \text{ not LS})_z$. In this expression, the ($Co^{3+}$ not LS) spin state is $Co^{3+}$ IS (with $z = 0.7$) in case of a pure intermediate state for trivalent cobalt, whereas it corresponds to $Co^{3+}$ HS (with $0 < z < 0.7$) within the framework of the mixed scenario. In our analysis, we also consider that $Co^{4+}$ can be either LS or IS, since the latter spin state can take place for tetravalent cobalt in an octahedral environment [18], although most of the previous studies only assumed a LS state.

Even if the VSST always takes place within the paramagnetic regime [3-17], a Curie-Weiss analysis of the susceptibility can only be performed within a temperature range located much above $T^*$, to ensure minimizing the impact of any temperature dependence of the spin-states themselves. On the other hand, a reliable evaluation of the Curie constant requires a broad enough temperature range. As a compromise, we will consider in the present study (where $T^* \sim 90$ K), a 150 K-wide temperature range centered at 300 K (i.e., 225-375 K).

## 2. Experimental details

Polycrystalline samples of $(Pr_{0.7}Sm_{0.3})_{0.7}Ca_{0.3}CoO_3$ were prepared by solid-state reaction using stoichiometric proportions of $Pr_6O_{11}$, $Sm_2O_3$, $CaO$ and $Co_3O_4$. To ensure good oxygen stoichiometry, the samples were first sintered at 1200 °C in flowing oxygen and then annealed in high-pressure (130 bar) $O_2$ atmosphere for 48 h at 600 °C. Powder x-ray diffraction showed these samples are monophasic, with orthorhombic symmetry (space group P*nma*) and parameters [$a = 5.3461(8)$ Å, $b = 7.5518(8)$ Å, and $c = 5.3499(7)$ Å] in line with the literature [6]. The magnetic measurements were recorded on a sample having a mass of ~28.3 mg and dimensions ~ $1.7 \times 1.7 \times 1.5$ mm$^3$ (the magnetic field being applied along the shortest dimension).

Magnetization data were recorded with a Superconducting Quantum Interference Device magnetometer (MPMS, Quantum Design), whose calibration was achieved using a Pd standard (cylinder of diameter ~ 2.9 mm and length ~ 3.2 mm) measured around 300 K in 10 kOe [19]. For $(Pr_{0.7}Sm_{0.3})_{0.7}Ca_{0.3}CoO_3$, an isofield magnetization curve, *M(T)*, was measured in 10 kOe upon warming from 10 K up to 400 K, while isothermal magnetization curves, *M(H)* were measured at a series of temperatures over this *T* range. All measurements were derived from 4 cm-length scans analysed with the iterative regression mode [20].

The temperature dependence of the dc susceptibility defined by $\chi = M/H$ was derived from the *M(T)* curve. One can consider that the uncertainty associated to such a measurement has three main contributions: $(\Delta\chi/\chi)_{mea.} = (\Delta\mathcal{M}/\mathcal{M}) + (\Delta m/m) + (\Delta H/H)$, where the first term refers to the magnetic moment, the second to the mass and the third to the measuring field. With our experimental device, the uncertainty in the absolute value of the magnetic moment is dominated by two contributions: (i) the uncertainty in the susceptibility of the calibration standard itself (± 0.5%) [19], and (ii) different deviations from the point-dipole approximation for the standard and for the measured sample [20]. For the Pd standard, the ratio between the longitudinal and transverse dimensions is 3.2/2.9 ~ 1.1, whereas it is 1.5/1.7 ~ 0.9 for our sample, introducing an error evaluated to be ~ 0.5% [21]. The uncertainty in mass is $\Delta m/m$ ~ 0.1/28.3 ~ 0.35 %, while that in field can be mainly ascribed to the presence of a remnant field trapped in the superconducting coil of the magnetometer (about 5 Oe after a degaussing procedure [20]) leading to $\Delta H/H$ ~ 0.05% in 10 kOe. Adding together these contributions leads to consider $(\Delta\chi/\chi)_{mea.}$ ~ 1.5 %.

## 3. Results and discussion

*3.1. Uncertainty affecting the χ(T) data*

In case of a magnetic susceptibility curve derived from *M(T)* data, the uncertainty to be considered is not only $(\Delta\chi/\chi)_{mea.}$; it must also reflect to which extent the ratio *M/H* is representative of the actual magnetic susceptibility. To address this issue, Figure 1 shows *M(H)* curves recorded at a few temperatures between 20 and 400 K. As previously reported in other compounds showing the VSST [3,5,9,22], *M(H)* curves at *T* << *T\** (presently exemplified by the 20 K curve) exhibit a downward curvature, most likely ascribable to the development of ferromagnetic interactions among the $Co^{4+}$ [3,9]. This feature progressively vanishes for *T* > *T\**, and the curves around room temperature appears to be perfectly linear in field.

To look at this issue in more detail, Fig. 2 displays three sets of *M(H)* data at 300 K, which correspond to independent series of measurements and which were recorded either upon increasing or decreasing the field. First, one observes that these curves are superimposed on each other, attesting to the good reproducibility of the data and the virtual absence of hysteretic effects. Second, one observes that *M* seems to be perfectly proportional to *H* in such *M(H)* curves. Nevertheless, a deviation from linearity can be revealed if one plots *M/H* versus *H*, as shown in the inset of Fig. 2. This feature is of importance for the present study, since it introduces a new source of uncertainty to be applied to the χ(*T*) curve derived from *M(T)* data recorded in a given field value (1 T in our case). The behavior of the apparent dc susceptibility (*M/H*), shown in the inset of Fig. 2, can be ascribed to the presence of a parasitic phase having a net magnetization; In this case, larger values of *M/H* in low fields would indeed result from the field-induced polarization of this magnetization, which then reaches saturation for fields larger than about 1 T [23]. The best way to eliminate such a contribution when attempting to determine the *true* susceptibility is to consider the derivative of *M(H)* in high fields. In the present study, we checked that *dM/dH* saturates in high fields, yielding the value shown by an horizontal line in the inset of Fig. 2, which is found to deviate from *M/H*(1T) by about 1%. The same features were observed over the whole *T* range between 225 and 375 K. Accordingly, beyond the uncertainty in the measurement itself [that we referred to as $(\Delta\chi/\chi)_{mea.}$], we suggest to incorporate a second term [denoted $(\Delta\chi/\chi)_{lin.}$] reflecting the fact that *M/H* is not perfectly field-independent. Writing the resulting net uncertainty as $(\Delta\chi/\chi) = (\Delta\chi/\chi)_{mea.} + (\Delta\chi/\chi)_{lin.}$, one obtains $(\Delta\chi/\chi) \sim 1.5 + 1 \sim 2.5\%$.

*3.2. Curie-Weiss analysis around RT*

Figure 3 shows the $\chi(T)$ curve derived from *M(T)* measured in 1 T. The VSST at $T^* \sim$ 90 K manifests itself by a sharp increase in susceptibility as the temperature is increased. The temperature range centered at 300 K that will be considered for the Curie-Weiss analysis is shown by the thick line. Ticks at each boundary of this range show the experimental uncertainty calculated from the above discussion (± 2.5%). To analyze the paramagnetic regime of $(Pr_{0.7}Sm_{0.3})_{0.7}Ca_{0.3}CoO_3$ in this temperature range, let us first specify the magnetic responses that are expected for each of the cations at play in this compound. Around room temperature (RT), experimental values of $\mu_{eff}$ for $Pr^{3+}$ are generally close to the free-ion one associated to the multiplet $^3H_4$ (3.58 $\mu_B$) [24,25]. We consider, in the present study, the value $\mu_{eff}(Pr^{3+}) = 3.60\mu_B$ experimentally derived both by Cohen *et al.* and Zhou *et al.* in the closely related perovskite $PrAlO_3$ [26,27]. For $Sm^{3+}$, the crystalline-electric-field has a stronger impact on magnetism than for $Pr^{3+}$, because of more closely spaced multiplets leading to *J*-mixing effects [24]. This generally yields a substantial temperature-independent paramagnetic (TIP) contribution, which can be predominant around RT [25,28]. In a detailed investigation of the perovskite $SmCoO_3$ ($Co^{3+}$ being LS), Ivanova *et al.* found that the paramagnetic response of $Sm^{3+}$ can be well described by $\chi(Sm^{3+}) = \chi_{TIP} + C/T$, with $\chi_{TIP}(Sm^{3+}) = 1.4 \ 10^{-3}$ emu/mol and $C = 0.0276$ emu.K/mol [29] which corresponds to $\mu_{eff}(Sm^{3+}) = 0.47\mu_B$. For both $Co^{4+}$ and $Co^{3+}$, we consider the spin-only values of $\mu_{eff}$, as usually done in the literature on cobaltites: $\mu_{eff}(Co^{4+}\ LS) = 1.73\mu_B$; $\mu_{eff}(Co^{4+}\ IS) = 3.87\mu_B$; $\mu_{eff}(Co^{3+}\ HS) = 4.90\mu_B$; $\mu_{eff}(Co^{3+}\ IS) = 2.83\mu_B$; and $\mu_{eff}(Co^{3+}\ LS) = 0\mu_B$. A TIP term exists for $Co^{3+}$ LS ($\approx$ 1.64 10$^{-4}$ emu/mol) [29,30], but it can be neglected in a first approximation owing to its smallness.

To isolate the Curie-Weiss part ($\chi^*$) of the experimental susceptibility, one subtracted the diamagnetic term as well as the TIP contribution from $Sm^{3+}$: $\chi^* = \chi - \chi_{dia} - 0.21\chi_{TIP}(Sm^{3+})$. Using $\chi_{dia}$ = -0.63 10$^{-4}$ emu/mol derived from tabulated values, and $\chi_{TIP}(Sm^{3+})$ = 1.4 10$^{-3}$ emu/mol, Fig. 4 shows the resulting $1/\chi^*$ vs. $T$ curve in a 150 K-wide temperature range around 300 K. A linear fitting of this curve yields values of the Curie-Weiss temperature and of the effective moment equal to ~ - 100 K and ~ 4.85 $\mu_B$, respectively. Are also displayed on Fig. 4 the error bars (at a few data points for the sake of clarity), considering $\Delta(1/\chi^*)/(1/\chi^*) \approx \Delta\chi/\chi \approx 2.5\%$ and $\Delta T/T \approx 0.5\%$. Since these errors can essentially be regarded as systematic errors, the uncertainty in the Curie constant derived from the slope of $1/\chi^*$ vs. $T$ is approximated by $\Delta C/C \approx \Delta(1/\chi^*)/\chi^* + \Delta T/T \approx 3\%$. With $\Delta\mu_{eff}/\mu_{eff} = (1/2)\Delta C/C \approx 1.5\%$, the experimental value of the effective moment is thus found to be $\mu_{eff}^{exp} = 4.85 \pm 0.07\mu_B$.

*3.3. The Co spin-states around RT*

Writing the cationic distribution as explained above, the experimental value $\mu_{eff}^{exp}$ must be compared to :

$$\mu_{eff}^{cal} = \sqrt{0.49[\mu_{eff}(Pr^{3+})]^2 + 0.21[\mu_{eff}(Sm^{3+})]^2 + 0.3[\mu_{eff}(Co^{4+})]^2 + z[\mu_{eff}(Co^{3+}\ not\ LS)]^2} \quad (1)$$

There are various possibilities for the spin states of $Co^{3+}$ and $Co^{4+}$. As previously noted, we consider that $Co^{3+}$ can be either IS ($z$ = 0.7) or a mixture LS/HS ($z$ being the amount of HS per f.u.), while $Co^{4+}$ can be either LS or IS. This leads to four different cases:

(1): [$Co^{4+}$ LS & $Co^{3+}$ IS] $\Rightarrow$ $\mu_{eff}^{cal} = 3.59\mu_B$

(2): [$Co^{4+}$ LS & $Co^{3+}$ (LS/HS)] : $\mu_{eff}^{cal}(z) = \mu_{eff}^{exp} \Rightarrow z = 0.68 \pm 0.03$

(3): [$Co^{4+}$ IS & $Co^{3+}$ IS] $\Rightarrow$ $\mu_{eff}^{cal} = 4.06\mu_B$

(4): [$Co^{4+}$ IS & $Co^{3+}$ (LS/HS)] : $\mu_{eff}^{cal}(z) = \mu_{eff}^{exp}$ $\Rightarrow z = 0.53 \pm 0.03$

Options (1) and (3) yield values of the effective magnetic moment that are significantly lower than the experimental one (4.85 ± 0.07 $\mu_B$), indicating that the occurrence of a pure IS state for $Co^{3+}$ can be discarded. One is thus led to favor the achievement of a mixed LS/HS state for $Co^{3+}$ at $T > T^*$, a behavior previously claimed to take place in $LaCoO_3$ for $T_1<T<T_2$ [31-37]. Option (2) leads to $z \sim 0.68$, which would mean the presence of a very small amount of $Co^{3+}$ LS (~ 2% per Co). This is hardly compatible with XAS data in which we observed that one of the hallmarks of $Co^{3+}$ LS (sharp feature at the low-energy side of the Co-$L_2$ edge) [35,38] remains well visible on the Co-$L_{2,3}$ spectra till RT [16]. In the end, the most likely option turns out to be (4), where $Co^{3+}$ is in a mixed LS/HS state, while $Co^{4+}$ is IS. It must be emphasized that this is consistent with the conclusion previously derived from the analysis of the XAS data [16].

Let us now evaluate the impact of two features that were neglected so far. First, XAS spectra exhibited a small peak close to 778 eV which can be ascribed to cobalt ions in a divalent state [16]. The fitting of these spectra led to a fraction of $Co^{2+}$ equal to ~ 2%, a parasitic contribution often found in cobalt perovskites [35], which may be related to a disproportionation process $2Co^{3+} \rightarrow Co^{2+} + Co^{4+}$ [39]. Denoting α the amount of $Co^{2+}$ per f.u., the cationic formulation must thus be rewritten as $(Pr^{3+})_{0.49}(Sm^{3+})_{0.21}(Ca^{2+})_{0.3}(Co^{2+})_\alpha(Co^{4+}IS)_{0.3+\alpha}(Co^{3+}LS)_{0.7-2\alpha-z}(Co^{3+}HS)_z$, leading the expected effective moment to be changed to $\sqrt{(\mu_{eff}^{cal})^2 + \alpha\{[\mu_{eff}(Co^{2+})]^2 + [\mu_{eff}(Co^{4+}IS)]^2\}}$. With $\mu_{eff}(Co^{2+}) = 3.87\mu_B$ and α = 0.02, this correction leads to decrease $z$ by ~ 0.03 with respect to the $Co^{2+}$-free evaluation. The second correction deals with the magnetic response of $Co^{3+}$ LS . If one takes into account $\chi_{TIP}(Co^{3+} LS) = 1.64 \cdot 10^{-4}$ emu/mol, the susceptibility to be fitted by a Curie-Weiss law becomes $\chi^{**} = \chi^* - (0.7-z)\times 1.64 \cdot 10^{-4}$. In this case, $z$ must be determined from the equation $\mu_{eff}^{exp}(z) = \mu_{eff}^{cal}(z)$, where $\mu_{eff}^{exp}(z)$ is the effective moment derived from a linear fitting of $1/\chi^{**}(z)$ vs. $T$ while $\mu_{eff}^{cal}(z)$ is given by equation (1). As anticipated, this correction as a very small impact, decreasing $z$ by only ~ 0.01. Accounting for these two corrections, our final estimate of the $Co^{3+}$ HS content around RT becomes $z =$ 0.49 ± 0.03. It is worth noticing that this value is in fairly good agreement with the result of the XAS analysis which led to $z \sim 0.4$ at $T = 290$ K [16].

*3.4. Comparison with the literature on $Pr_{0.5}Ca_{0.5}CoO_3$*

Let us now comment on results of the literature about $Pr_{0.5}Ca_{0.5}CoO_3$, which is the archetypical material exhibiting a VSST. In a recent XAS study, Herrero-Martín *et al.* [15] reported that the $Co^{3+}$ spin state around RT ($>> T^* \sim 75$ K) can be equally well described by a pure IS state or a mixed LS/HS state. In the latter case, the ratio LS:HS was estimated to be 50:50, which would correspond to a content of $Co^{3+}$ HS per f.u. equal to $z = 0.25$. Besides, Barón-González *et al.* [10] deduced for this compound $\mu_{eff}^{exp}(Co) = 3.64\mu_B$ from the Curie-Weiss behavior observed in the range ~ 150-350 K. Following the same approach as that described above, this $\mu_{eff}^{exp}(Co)$ value should be compared to $\mu_{eff}^{cal}(Co) = \sqrt{0.5[\mu_{eff}(Co^{4+})]^2 + 0.5[\mu_{eff}(Co^{3+})]^2}$. If ones assumes $Co^{4+}$ to be in a LS state, $\mu_{eff}^{exp}(Co) = 3.64\mu_B$ requires that $\mu_{eff}(Co^{3+}) = 4.84\,\mu_B$. It turns out that this value is much larger than that of $Co^{3+}$ IS (2.83 $\mu_B$), while, in the frame of a LS/HS coexistence, it would mean that ~98 % of the $Co^{3+}$ are HS, which is at odds with the XAS data [15]. Alternatively, considering $Co^{4+}$ in an IS state would imply that $\mu_{eff}(Co^{3+}) = 3.39\,\mu_B$. One the one hand, this value is again incompatible with $Co^{3+}$ IS, but on the other hand, it corresponds to a mixture LS/HS with $z = 0.24$, a value which turns out to be remarkably close to that estimated from XAS ($z = 0.25$) [15]. Therefore, these data on $Pr_{0.5}Ca_{0.5}CoO_3$ are found to be well consistent with our conclusions derived from $(Pr_{0.7}Sm_{0.3})_{0.7}Ca_{0.3}CoO_3$, namely the occurrence of $Co^{4+}$ in an IS state and of $Co^{3+}$ in a mixed LS/HS state above the VSST.

## 4. Conclusion

$(Pr_{0.7}Sm_{0.3})_{0.7}Ca_{0.3}CoO_3$ exhibits a valence and spin state transition (VSST) at $T^* \approx 90$ K. We performed a quantitative analysis of the magnetic susceptibility around room temperature, i.e. at $T >> T^*$, in a simple ionic picture. In agreement with a previous XAS experiment, it was found that $Co^{3+}$ is in a mixed state combining LS and HS spin-states, while the spin state of $Co^{4+}$ is most likely IS. We also noted that this configuration of Co spin-states above the VSST is consistent with data of the literature on $Pr_{0.5}Ca_{0.5}CoO_3$. More quantitatively, the population of $Co^{3+}$ HS around 300 K was evaluated to be ~ 0.49 per f.u. for $(Pr_{0.7}Sm_{0.3})_{0.7}Ca_{0.3}CoO_3$, which is in reasonable agreement with the value previously derived from XAS (~ 0.4 per f.u.). Similar analysis on other compounds showing the VSST would be useful to further test the generality of these findings.

## Figure captions

Fig.1: Isothermal magnetization curves at selected temperatures (measured by increasing the magnetic field after a zero-field cooling).

Fig.2: Three magnetization curves at 300 K, measured either in increasing (filled circles and up triangles) or decreasing (down triangles) magnetic field. The inset shows the field dependence of the ratio *M/H* for each of these curves. The horizontal line is the value towards which saturates the derivative *dM/dH* in high fields (i.e., above ~ 3 T).

Fig. 3: Temperature dependence of the dc susceptibility $\chi$ (= *M/H*) derived from *M(T)* recorded in 1 T. The thick line highlights the temperature range used for the Curie-Weiss analysis. The ticks at selected temperatures correspond to the net uncertainty attributed to this $\chi(T)$ data (see text).

Fig. 4: Temperature dependence of the inverse of the Curie-Weiss part of the magnetic susceptibility ($\chi^*$), obtained by subtracting from $\chi$ the diamagnetic susceptibility and the temperature-independent paramagnetic term associated to $Sm^{3+}$. The ticks mark the experimental uncertainties that were considered for $1/\chi^*$ ($\pm$ 2.5%) and for *T* ($\pm$ 0.5%).

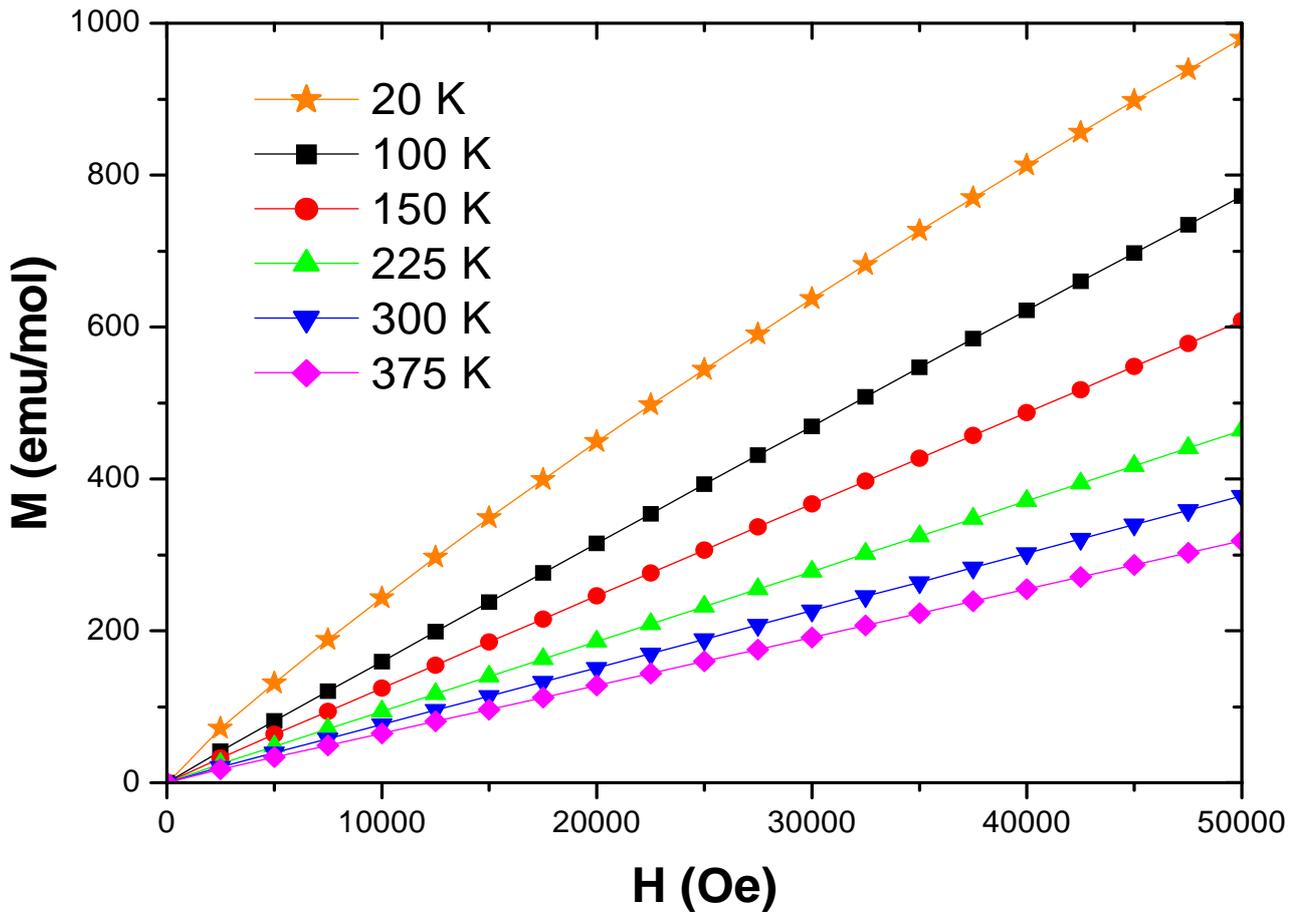

Figure 1
Guillou et al.

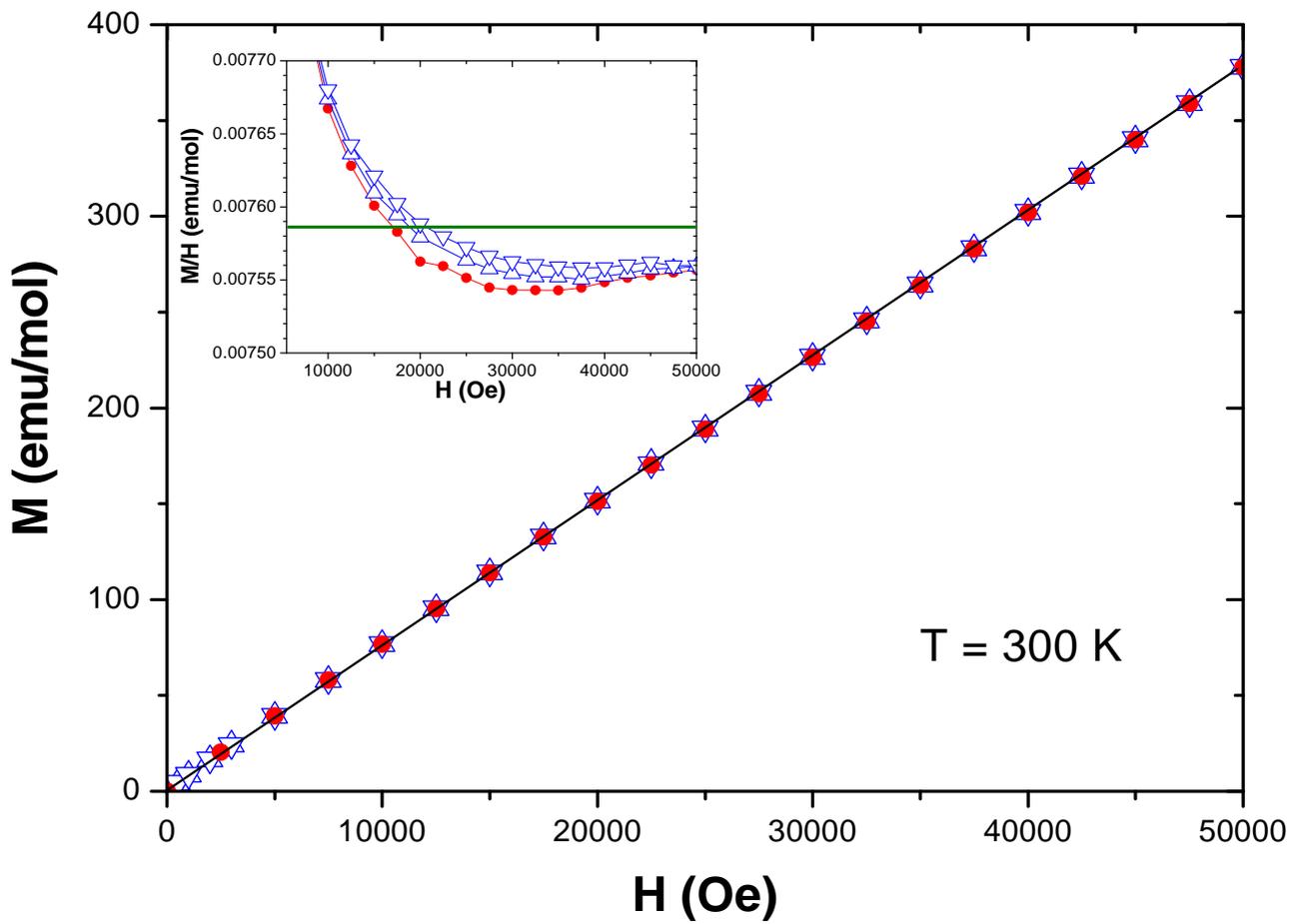

Figure 2
Guillou et al.

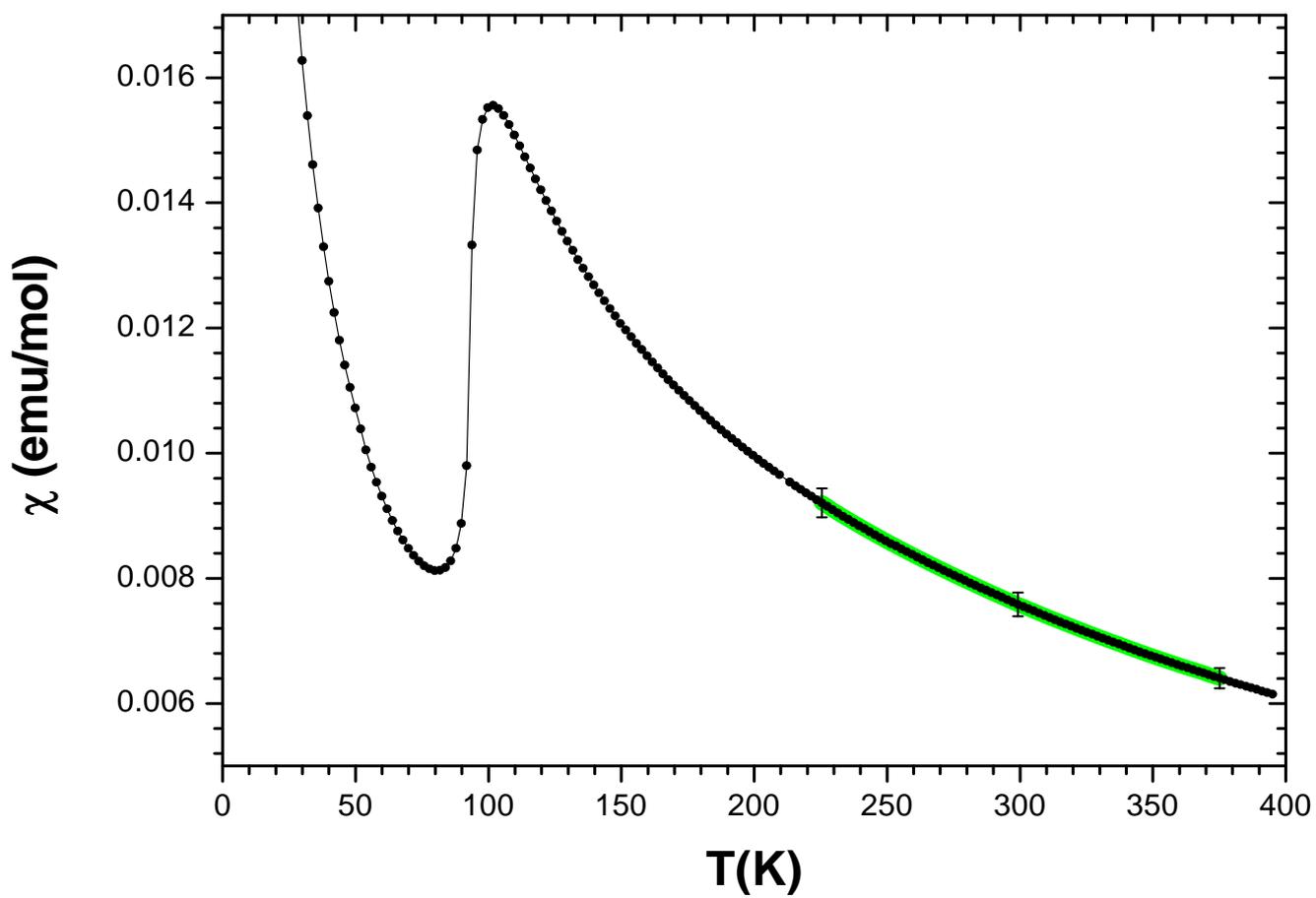

Figure 3
Guillou et al.

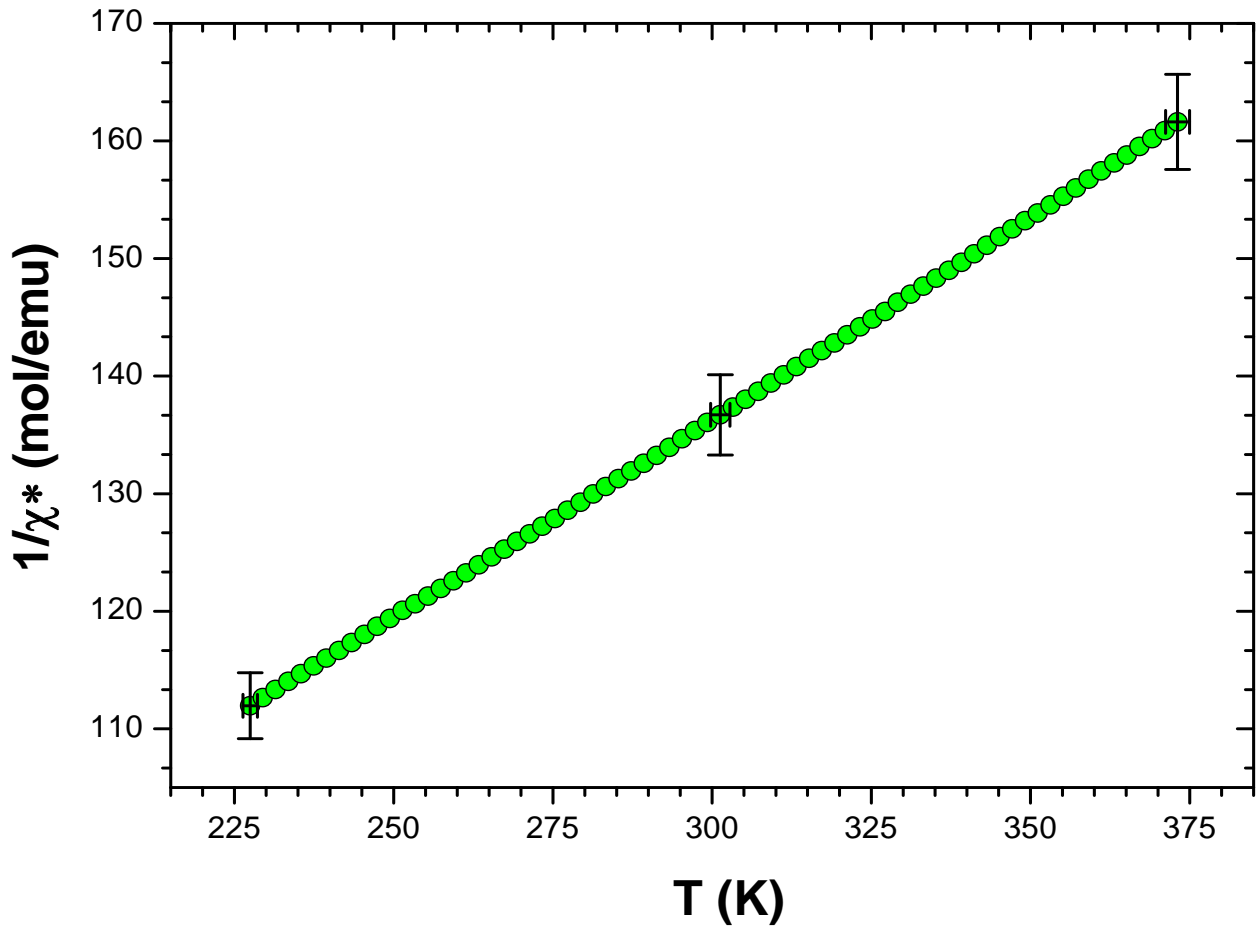

Figure 4
Guillou et al.